\documentclass[trackchanges]{aastex7}

\usepackage{footnote}
\usepackage{amsmath}
\def\oc3{[O~{\sc iii}]$\lambda5007$\AA}
\def\oee{[O~{\sc iii}]$_e\lambda5003$\AA}
\def\nii{[N~{\sc ii}]$\lambda6583$\AA}
\def\nei{[N~{\sc ii}]$_e\lambda6583$\AA}
\def\si{[S~{\sc ii}]$\lambda6716$\AA}
\def\oi{[O~{\sc i}]$\lambda6300$\AA}
\def\oii{[O~{\sc i}]$\lambda6363$\AA}
\def\sii{[S~{\sc ii}]$\lambda6732$\AA}
\usepackage{lineno}

\usepackage[caption=false]{subfig}
\usepackage{multirow}
\usepackage{graphicx}
\usepackage{booktabs}
\usepackage{amsmath,amssymb}
\usepackage[T1]{fontenc}
\usepackage{booktabs}
\usepackage{threeparttable}

\begin{document}
\nolinenumbers
\title{A candidate for True Type-2 AGN without hidden central BLRs Identified by central Tidal Disruption Event}

\correspondingauthor{Xue-Guang Zhang; En-Wei Liang}
\email{xgzhang@gxu.edu.cn; lew@gxu.edu.cn }

\author[0009-0005-6943-7803]{Ying Gu}
\affiliation{Guangxi Key Laboratory for Relativistic Astrophysics, School of Physical Science and Technology, \\
Guangxi University, Nanning 530004, People's Republic of China}
\email{going@st.gxu.edu.cn} 

\author[0000-0003-4583-3383]{Qi Zheng}
\affiliation{School of Physics and technology, Nanjing Normal University, No. 1, Wenyuan Road, Nanjing,
210023, People's Republic of China}
\email{zhengqi@nnu.edu.cn} 

\author[0009-0004-9177-2934]{Pei-zheng Cheng}
\affiliation{Guangxi Key Laboratory for Relativistic Astrophysics, School of Physical Science and Technology, \\
Guangxi University, Nanning 530004, People's Republic of China}
\email{2307301016@st.gxu.edu.cn} 

\author[0009-0001-5037-3855]{Xiao Li}
\affiliation{Guangxi Key Laboratory for Relativistic Astrophysics, School of Physical Science and Technology, \\
Guangxi University, Nanning 530004, People's Republic of China}
\email{lxgo@st.gxu.edu.cn}

\author[0009-0005-0082-2017]{Xing-Qian Cheng}
\affiliation{Guangxi Key Laboratory for Relativistic Astrophysics, School of Physical Science and Technology, \\
Guangxi University, Nanning 530004, People's Republic of China}
\email{2307301014@st.gxu.edu.cn}

\author[0000-0001-5458-8354]{Xue-Guang Zhang$^*$}
\affiliation{Guangxi Key Laboratory for Relativistic Astrophysics, School of Physical Science and Technology, \\
Guangxi University, Nanning 530004, People's Republic of China}
\email{xgzhang@gxu.edu.cn}

\author[0000-0002-7044-733X]{En-Wei Liang$^*$}
\affiliation{Guangxi Key Laboratory for Relativistic Astrophysics, School of Physical Science and Technology, \\
Guangxi University, Nanning 530004, People's Republic of China}
\email{lew@gxu.edu.cn}

\begin{abstract}

In this manuscript, through applications of TDE (tidal disruption event) expected variability properties, a potential 
candidate for True type-2 AGN without hidden central broad line regions (=TT2AGN) is reported in the SDSS J233454.07+145712.9 
(=SDSS J2334). Through analyzing the 20-years optical light curves of SDSS J2334 from different Sky Survey projects, a TDE is 
preferred with a $4.7{\rm M_\odot}$ main-sequence star tidally disrupted by the central BH with mass $11.7\times 10^6{\rm M_\odot}$, 
indicating that central region within distance about 20 light-days to central BH in SDSS J2334 is directly in the line-of-sight. 
Moreover, AGN activities in SDSS J2334 can be confirmed through applications of BPT diagrams. Meanwhile, comparing virial BH 
mass determined through assumed broad Balmer emission components and M-sigma expected BH mass by well measured stellar velocity 
dispersion through stellar absorption features, optical broad emission lines in SDSS J2334 are disfavored with confidence level 
higher than 6$\sigma$. Therefore, combining the unique properties of the TDE and the spectroscopic results with only narrow 
emission lines, SDSS J2334 can be well identified as a potential candidate for a TT2AGN. The results indicate the to detect 
TDE expected flares in normal Type-2 AGN  classified by spectroscopic results should be a new practicable method for identifying 
TT2AGN.

\end{abstract}


\section{Introduction} \label{sec:intro}
	Active galactic nuclei (AGNs) are broadly classified into two different types within the Unified Model (UM), i.e., Type 1 
AGNs have broad line regions (BLRs) and narrow-line regions (NLRs) in the line of sight, while Type 2 AGNs are obscured by a dusty 
torus, with only the NLRs observed \citep{1993ARA&A..31..473A,1995PASP..107..803U}. Although the UM successfully explains broad 
observations, some additional ingredients are needed in order to take into account all the observational evidence (e.g. 
\citet{1997ApJ...477..631V,2001ApJ...554L..19T,2005AJ....129.1212Z, 2005ApJ...633..638W,2009AJ....137.5120Y,
2018A&A...614A.120S}). More recent detailed discussions on the UM can be found in \citet{2015ARA&A..53..365N}. Even after considering 
necessary modifications, the UM is an incomplete picture for accommodating the data (e.g., \citet{2006ApJ...653L..25G, 
2012ApJ...747L..33E}). As proposed by \citet{2001ApJ...554L..19T,2003ApJ...583..632T}, a significant fraction of Type 2 AGN does 
not show hidden BLRs, even when high-quality spectropolarimetric data are available, leading to the called True Type-2 AGN (TT2AGN) 
in the manuscript. Here, TT2AGN, also called Type-2 AGN with non-hidden BLRs or AGN without BLRs or unobscured 
Type-2 AGN or naked AGN in the literature, means the AGN which has no central BLRs, not the AGN with heavily obscured central BLRs 
leading to the classified Type 1.9 AGN. Since then, unobscured narrow-line AGN were well studied for identifying TT2AGN without 
hidden central BLRs as predicted by the UM from both the observations \citep{2004A&A...424..519H,2008MNRAS.390.1241B, 
2011ApJ...734L..16H,2012MNRAS.426.3225B,2012ApJ...759L..16H,2012A&A...544A.129V,2014MNRAS.438..557Z,2014AJ....147...12B,
2015ApJ...803...57I,2016A&A...594A..72P,2021ApJ...922..248X} and theoretical analysis \citep{2009ApJ...701L..91E, 
2010ApJ...724..855C,2015ApJ...803...57I,2015AJ....149...75L,2016MNRAS.459..585E}. However, the existence of TT2AGN is still under 
debating \citep{2004AJ....128.2066N,2012A&AT...27..557A,2015ApJ...803...57I}. For example, observations with HST suggest that 
the previously accepted TT2AGN candidate, NGC 3147, is definitely not a TT2AGN \citep{2019MNRAS.488L...1B}. Therefore, 
identifying candidates of TT2AGN remains extremely challenging.

	Based on the UM expected obscurations on central AGN activity in normal type-2 AGN, there should be no photometric 
variability in normal Type-2 AGN. Therefore, an effective approach to identify a TT2AGN candidate is to directly search for 
significant flux variability of normal Type-2 AGN \citep{2004A&A...424..519H,2014AJ....147...12B}. Here, Tidal disruption events 
(TDEs) can produce a bright flare that may last for months to years (e.g., \citealt{1988Natur.333..523R, 2015MNRAS.452...69M, 
2019ApJ...883...94T,2022MNRAS.516L..66Z}). These flares occur when a star's orbit passes within the tidal radius of a supermassive 
black hole (SMBH), leading TDEs as reliable indicators of BH accreting systems located within dozens to hundreds of Schwarzschild 
radii from central SMBHs \citep{1988Natur.333..523R,2011Natur.476..421B, 2012Natur.485..217G,2016Sci...351...62V, 
2018Sci...361..482M,2021ARA&A..59...21G}. For example, the well-known TDE ASASSN-14li has bright flares, lasting for six months 
\citep{2017MNRAS.466.4904B}, and has been extensively studied in recent years\citep{2024ApJ...963...75W,2024arXiv241112247Z}. 
Several samples of TDE candidates have been identified \citep{2021ApJ...908....4V,2021MNRAS.508.3820S,2023ApJ...942....9H, 
2023ApJ...955L...6Y}. A TDE observed in a normal Type 2 AGN classified by spectroscopic properties should be a smoking-gun of a 
TT2AGN. 

In this manuscript, considering the unusual outburst events in SDSS J233454.07+145712.9 (hereafter SDSS J2334) which has been reported in \citet{2019MNRAS.482...98D}, combining with spectroscopic emission line properties. We report the discovery of a TDE candidate in the type-2 AGN SDSS J2334, providing interesting clues to support SDSS J2334 as a potential candidate for 
TT2AGN. The manuscript is organized as follows. Section 2 presents the TDE candidate discovered from the optical long-term photometric light curves of SDSS J2334. Section 3 presents our spectroscopic analysis for SDSS J2334. Conclusions are presented in Section 4. The cosmological parameters of $H_{0}$=70 km s$^{-1}$ Mpc$^{-1}$, $\Omega_{m}$=0.3, and $\Omega_{\Lambda}$=0.7 are adopted.

\section{A TDE candidate in the long-term light curves of SDSS J2334} \label{sec:Light_Curve}
\begin{figure*}
\centering
\includegraphics[width = 18cm, height=12cm]{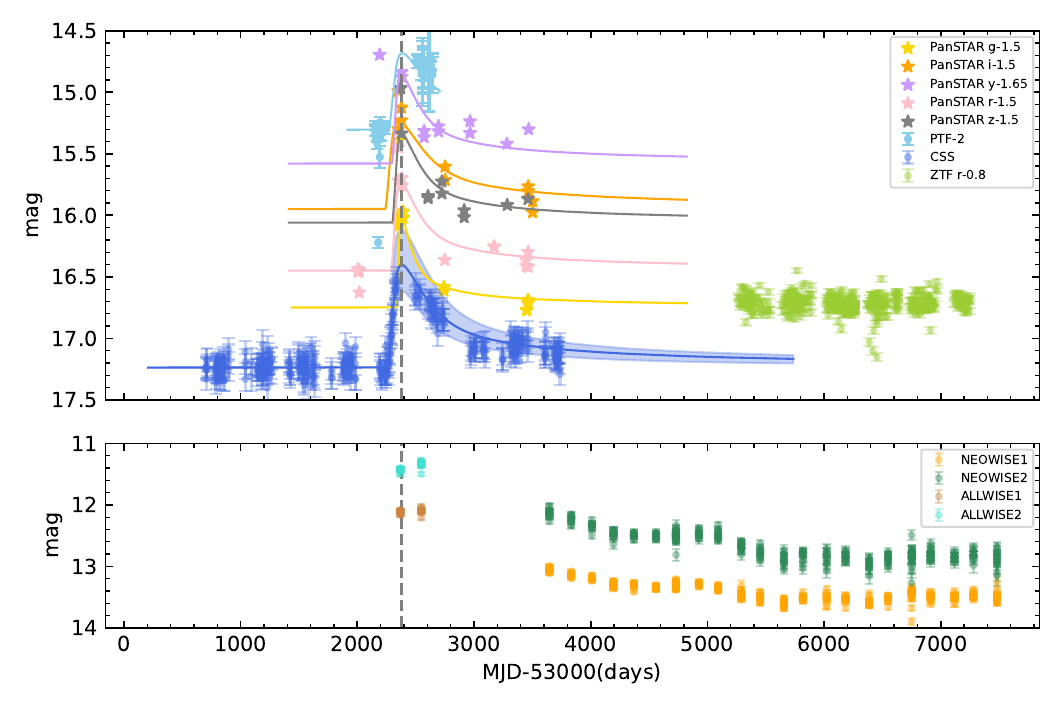}
\includegraphics[width = 18cm, height=4cm]{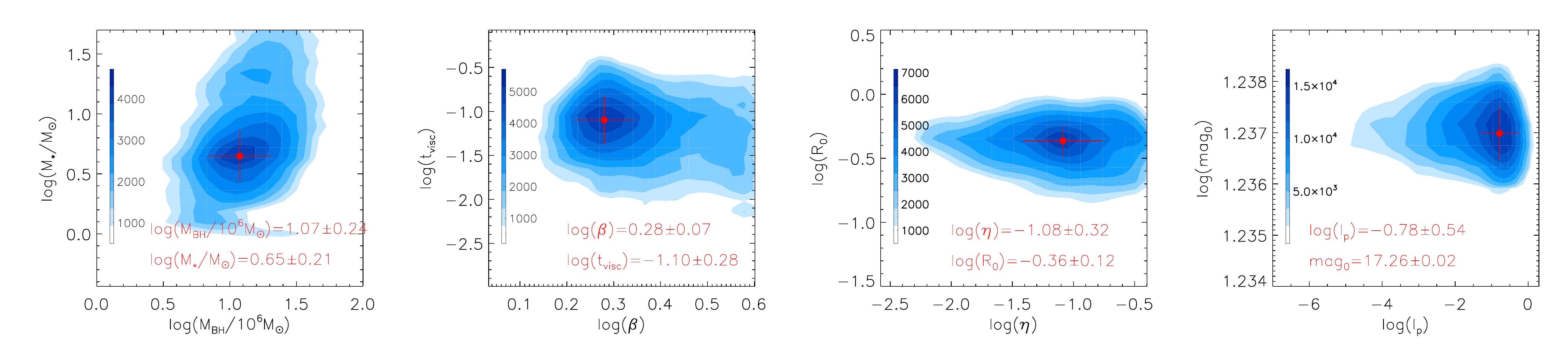}
\caption{The top panel shows the CSS V-band light curve (blue symbols) and the TDE model determined best 
descriptions (solid blue line) and the corresponding confidence bands (blue shaded area) determined by uncertainties of the 
model parameters. The remaining light curves are collected from the other projects, as shown in the legend in the top right 
corner. Solid lines in different colors show the best descriptions to the light curves after accepting the TDE model 
parameters applied to describe the CSS V band light curve, except for the ZTF r-band light curve. The middle panel displays 
the WISE infrared light curves of SDSS J2334 in the W1 and W2 bands, with different colors representing different observational 
data (see top-right legend). In the top two panels, the grey dashed line shows the peak positions preferred by the standard 
TDE model. Bottom panels show the MCMC determined two-dimensional distributions of TDE model parameters. In each bottom 
panel, solid circle plus error bars in red mark the accepted values and the corresponding uncertainties of the model 
parameters which are marked in red characters in bottom region.}
\label{LC}
\end{figure*}

	SDSS J2334 at redshift $z=0.1075$ has its 18years-long light curves shown in Fig.~\ref{LC}, including the Catalina Sky Survey 
(CSS) \citep{2009ApJ...696..870D} $V$ band light curve with MJD from 53554 (Jul. 2005) to 56587 (Oct. 2013), the Palomar Transient 
Factory (PTF) \citep{2009PASP..121.1395L} $R$ band light curve with MJD from 55023 (Jul. 2009) to 55499 (Oct. 2010), the PanSTARRS 
(Panoramic Survey Telescope \& Rapid Response System) $grizy$ band light curves with MJD from 55092 (Sep. 2009) to 56651 (Dec. 2013).
and the Zwicky Transient Facility (ZTF \citet{2019PASP..131a8002B}) r band light curve with MJD from 58268 (May 2018) to 60245 (Oct. 2023)\footnote{The ZTF-$g/i$ band light curves were not collected, as their variability is similar to that of the ZTF-$r$ band.} Through the shown light curves, two interesting points can be well confirmed. First, there is an outburst event with MJD-53000 from about 2100 (Oct. 2009) to about 3500 (Sep. 2013), not only similar as previously reported results in 
\citet{2019MNRAS.482...98D} through the CRTS (Catalina Real-Time Survey) V-band light curve, but also re-confirmed by the PanSTARRS 
and PTF light curves. Second, except the outburst event, SDSS J2334 is at quiescent state with none apparent variabilities, not 
only confirmed by the CSS V-band light curve with MJD-53000 smaller than 2000, but also confirmed by the more recent ZTF r-band 
light curves. This is further supported by ZTF g-band forced photometry (MJD 58290-60694) \citep{2023arXiv230516279M}, 
which shows no post-outburst variability (Appendix Fig A).

	For the outburst event with MJD-53000 from 2100 to 3500, there is a steep rise and then followed by a smoothly declined 
trend, similar as optical TDEs expected variability patterns. To reveal the nature of the outburst, we further describe the light 
curves with the conventional theoretical TDE model. We adopt the public codes of TDEFIT and MOSFIT \citep{2014ApJ...783...23G, 
2018ApJS..236....6G, 2019ApJ...872..151M} on theoretical TDE model. Similar as what we have recently done in 
\citet{2024MNRAS.tmp....2G,2024MNRAS.534L..23Z, 2025arXiv250116585Z} considering theoretical TDE model expected time dependent 
templates of viscous delayed accretion rates applied in simple black-body photosphere model, we outline the following procedure 
to describe the optical outburst event by theoretical TDE model.

	First, the viscous-delayed accretion rates $\dot{M}_{a}$ are created based on the $dM/dE$ (the fundamental elements in 
the public code of TDEFIT/MOSFIT) and the viscous delay effects \citep{2013ApJ...767...25G, 2019ApJ...872..151M}. A standard TDE 
event, in which a main-sequence star with $M_{*}=1{\rm M_\odot}$ is tidally disrupted by a BH with $M_{BH}$= $10^{6}{\rm M}_\odot$, 
is considered. We construct the standard templates of the time-dependent viscous-delayed accretion rate $\dot{M}_{a}(T_{v},\beta)$, 
where $\beta$ is the impact parameter and $T_{v}$ is the viscous time.

	Second, for common TDEs, the actual accretion rate at time $t$ can be estimated using scaling relations, which depend 
on the black hole mass and the stellar parameters \citep{2019ApJ...872..151M}:
\begin{equation}
	\begin{split}
		&\dot{M} = M_{\rm BH,6}^{-0.5}\times M_{*}^2\times
		R_{*}^{-1.5}\times\dot{M}_{a}(T_{v}, \beta) \\
		&t = (1+z)\times M_{\rm BH,6}^{0.5}\times M_{*}^{-1}\times
		R_{*}^{1.5} \times t_{a}(T_{v}, \beta),
	\end{split}
\end{equation}
where central BH mass $M_{\rm BH,6}$ is in units of ${\rm 10^6M_\odot}$, stellar mass $M_{*}$ is in units of ${\rm M_\odot}$, 
stellar radius $R_{*}$ is in units of ${\rm R_{\odot}}$, $z$ represents redshift of the host galaxy of the TDE. Additionally, 
we adopt the known mass-radius relation for main-sequence stars from \cite{1996MNRAS.281..257T}.

	Third, based on the standard blackbody photosphere (e.g., \citealt{2019ApJ...872..151M}), the time-dependent emission 
spectrum in the rest frame can be calculated as
\begin{equation}
	\begin{split}
		\begin{aligned}
			&T_p(t) = \left (\frac{L}{4\pi\sigma_{SB}R_p^2}\right)^{1/4} =
			\left[ \frac{\eta\dot{M(t)}c^2}{4\pi\sigma_{SB}R_p^2}\right ]^{1/4},\\
			&R_p(t) = R_0\times a_p(L/L_{\rm Edd})^{l_p} = R_0\times a_p\left [\frac{\eta\dot{M(t)}c^2}{1.3\times10^{38}M_{\rm BH}}\right ]^{l_p},\\
			&a_p = \left [G M_{\rm BH}\times (\frac{t_p}{2\pi})^2\right]^{1/3},	\\
			&F_\lambda(t)=\frac{2\pi hc^2}{\lambda^5}\frac{1}{e^{hc/(k\lambda T_p(t))}-1}\left[\frac{R_p(t)}{D(z)}\right]^2,\\
		\end{aligned}
	\end{split}
\end{equation}
where $T_p(t)$ and $R_p(t)$ are the time-dependent effective temperature and radius of the photosphere, respectively. $R_p(t)$ ranges 
from the minimum $R_{\rm isco}$ (event horizon radius) to $a_p$ (maximum semimajor axis) of the accreting mass. $L$ is the 
time-dependent bolometric luminosity given by $L = {\eta\dot{M(t)}c^2}$, $L_{\rm Edd}$ represents the Eddington luminosity, $\eta$ 
represents the energy transfer efficiency that is less than 0.4 \citep{2014ApJ...783...23G, 2019ApJ...872..151M}, $\sigma_{\rm SB}$ 
is the Stefan-Boltzmann constant, $l_p$ represents the power-law exponent and $t_p$ is the time information of the peak accretion. 
$D(z)$ is the luminosity distance at redshift $z$. Then, time-dependent apparent CSS-$V$ band magnitudes can be well determined through 
the $F_\lambda(t)$ in the observer frame convolving with the accepted transmission curves of the CSS-$V$ band filter.

	The free parameters of the TDE model include $M_{\rm BH}$, $M_{*}$ (the corresponding stellar radius $R_{*}$ calculated by the 
mass-radius relation), $\beta$, $T_{v}$, $\eta$, $R_0$, $l_p$, and the magnitude of the host galaxy in the CSS-$V$ band ($mag_0$). 
Then, through the known Maximum likelihood method combined with the Markov Chain Monte Carlo (MCMC) technique 
\citep{2013PASP..125..306F}, the best descriptions to the CSS V band light curve of SDSS J2334 can be determined and shown in the 
top panel of Fig.~\ref{LC}. Meanwhile, the MCMC technique determined posterior distributions of the model 
parameters are shown in the bottom panel of Fig.~\ref{LC}, leading the determined model parameters with a polytropic index of 
$\gamma=4/3$ to be: $\log(M_{\rm BH,~6})~\sim~1.07\pm0.24$, $\log(M_{*}/M_\odot)~\sim~0.65\pm0.21$, $\log(\beta)~\sim~0.28\pm0.07$, 
$\log(T_{v})~\sim~-1.10\pm0.28$, $\log(\eta)~\sim~-1.08\pm0.32$, $\log(R_{0})~\sim~-0.36\pm0.12$, $\log(l_{p})~\sim~-0.78\pm0.54$ 
and $mag_0~\sim~17.26\pm0.02$. Furthermore, by applying the obtained fitting parameters to the light curves in the PTF-$R$ and 
PanSTARRS-$grizy$ bands (the host galaxy magnitude in each band is treated as a free parameter), reasonable fitting results are 
also obtained, as shown in the middle panel of Fig \ref{LC}.

	Based on the theoretical TDE model determined results, the expected tidal disruption radius is about 
7$R_G$ (Schwarzschild radius). Once the central TDE can be expected, spatial size of the geometric structure of TDE debris can be 
simply checked. The TDE debris in SDSS J2334 around MJD-53000=3000~5000 have outer distance \citep{2014ApJ...783...23G}
\begin{equation}
	R_{out}~\sim~2\times(\frac{GM_{BH}t^2}{\pi^2})^{1/3}~\sim~7.6~-~16.6{\rm light-days}
\end{equation}
with $G$ as gravitational constant and $M_{BH}$ as the BH mass in units of ${\rm M_{\odot}}$, clearly indicating the central region 
around the central BH of SDSS J2334 is in the direction of our sight. 

To conclude this section, we note that it seems that a `bump' flare has been observed about $\sim$ 900 days after the
peak. Such rebrightening bump phenomena in optical band have been observed in some TDEs and candidates such as SDSS J0159+0033 
\citep{2015MNRAS.452...69M}, IRAS F01004-2237 \citep{2024A&A...692A.262S}, and some recent TDEs reported in \citet{2023ApJ...955L...6Y}. 
Several physical mechanisms have been proposed to explain the rebrightening bumps in TDEs, such as the transition from super-Eddington 
to sub-Eddington accretion \citep{2009MNRAS.400.2070S} and delayed disk formation \citep{2022ApJ...925...67L}. Recently, \citet{2025ApJ...979..235G} have collected a small sample of TDEs exhibiting prominent rebrightening bumps and have found that the model combining stream collisions and delayed accretion disk formation could provide a unified explanation for diverse TDE observations. We should note that the `bump' in SDSS J2334 falls within the confidence interval, but its character is less prominent compared to the aforementioned TDEs. Therefore, we only consider theoretical models of TDEs in our analysis.  As these discussions are beyond the scope of this manuscript, we will not explore them further.

\section{Spectroscopic results of SDSS J2334}\label{sec: spec}

\begin{figure*}
\centering\includegraphics[width=18cm,height=8cm]{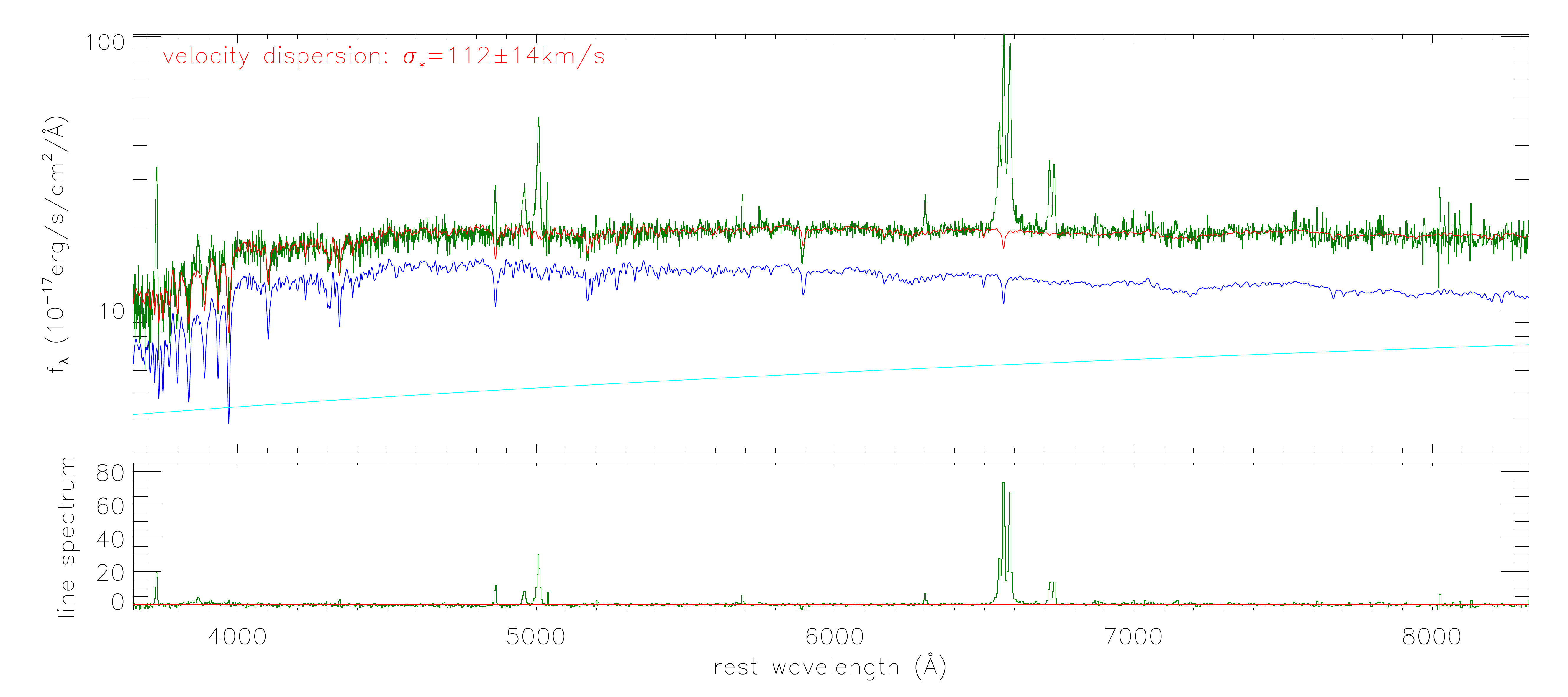}
\centering\includegraphics[width=18cm,height=8cm]{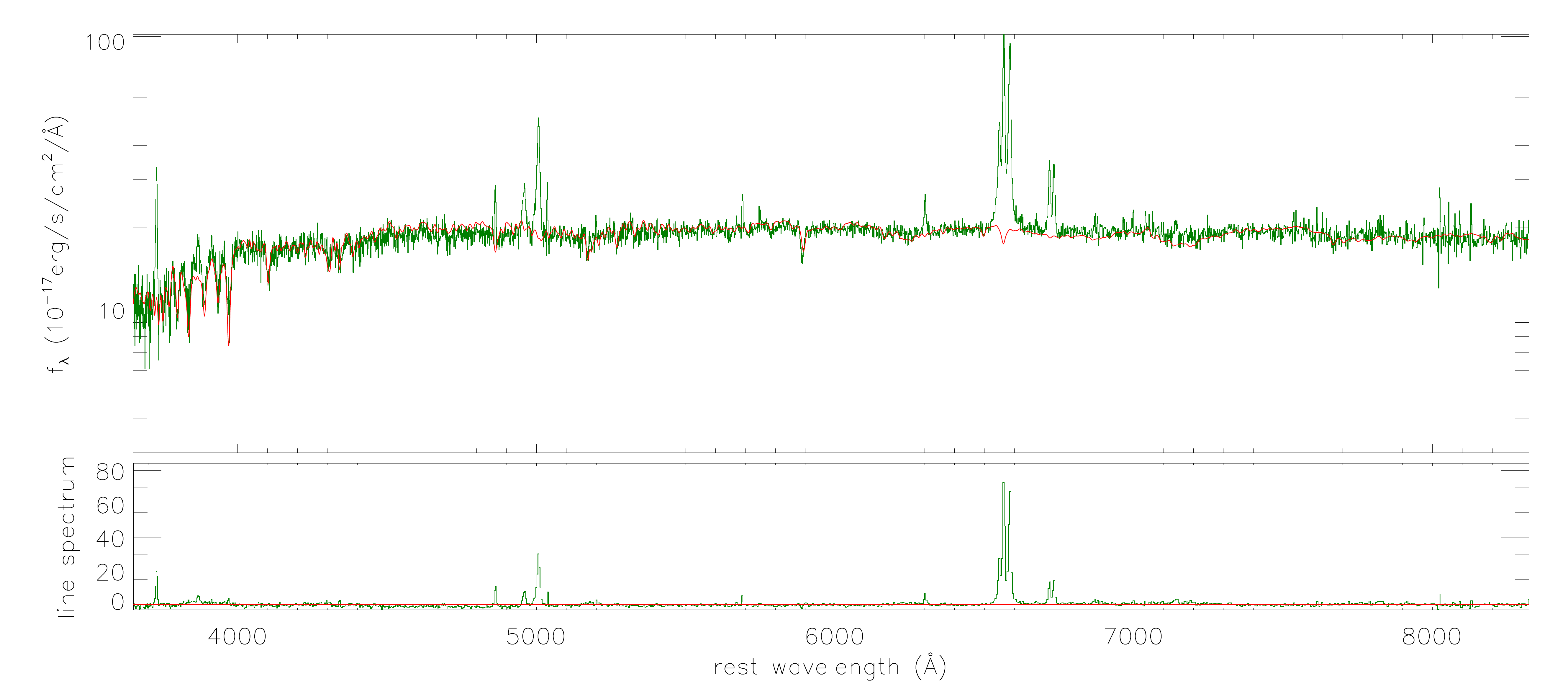}
\caption{Top panel shows the SDSS spectrum (in dark green) of SDSS J2334 and the SSP method determined best descriptions (in red) 
including the host galaxy contributions (in blue) and the AGN continuum emissions (in cyan), and the line spectrum in bottom region 
calculated by SDSS spectrum minus the host galaxy contributions plus the AGN continuum emissions. Bottom panel shows the corresponding 
results but without considerations of AGN continuum emissions.}
\label{spec}
\end{figure*}

	SDSS J2334 has its SDSS spectrum (median signal-to-noise about 18, plate-mjd-fiberid=0747-52234-0400) observed in MJD=52234 
(Nov. 2001, far away from the duration of the expected central TDE) collected from the Sloan Digital Sky Survey (SDSS) DR16 
\citep{2020ApJS..249....3A} and shown in top panel of Fig.~\ref{spec}, with apparent stellar absorption features. In order to 
measure emission lines, the well accepted SSP (Simple Stellar Populations) method \citep{1993ApJ...405..538B,2003MNRAS.344.1000B,
2005MNRAS.358..363C,2017MNRAS.466..798C} is applied to determine host galaxy contributions and intrinsic AGN continuum emissions, 
considering the 39 simple stellar population templates \citep{2003MNRAS.346.1055K,2003MNRAS.344.1000B} for host galaxy contributions 
plus a power law component for probably intrinsic AGN continuum emissions. The SSP method determined best descriptions to the SDSS 
spectrum with emission lines being masked out are shown in top panel of Fig.~\ref{spec} with  $\chi^2/dof\sim1.09$, through the 
well applied Levenberg-Marquardt least-squares minimization technique. And the corresponding line spectrum is shown in bottom region 
of panel a in Fig.~\ref{spec}.

	Meanwhile, similar fitting procedure was applied to describe the SDSS spectrum with emission lines being masked out in SDSS 
J2334, only considering the 39 simple stellar population templates but not considering continuum emission component, leading the 
determined descriptions shown in bottom panel of Fig.~\ref{spec} with corresponding $\chi^2/dof~\sim~1.36$, through the 
Levenberg-Marquardt least-squares minimization technique. The known F-test technique is applied to confirm confidence level 
higher than 10$\sigma$ to support the existence of the power law continuum emission component, after comparing with from different 
model functions with 2 as the difference between the different $dofs$ .

\begin{figure}
\centering\includegraphics[width=5cm,height=6cm]{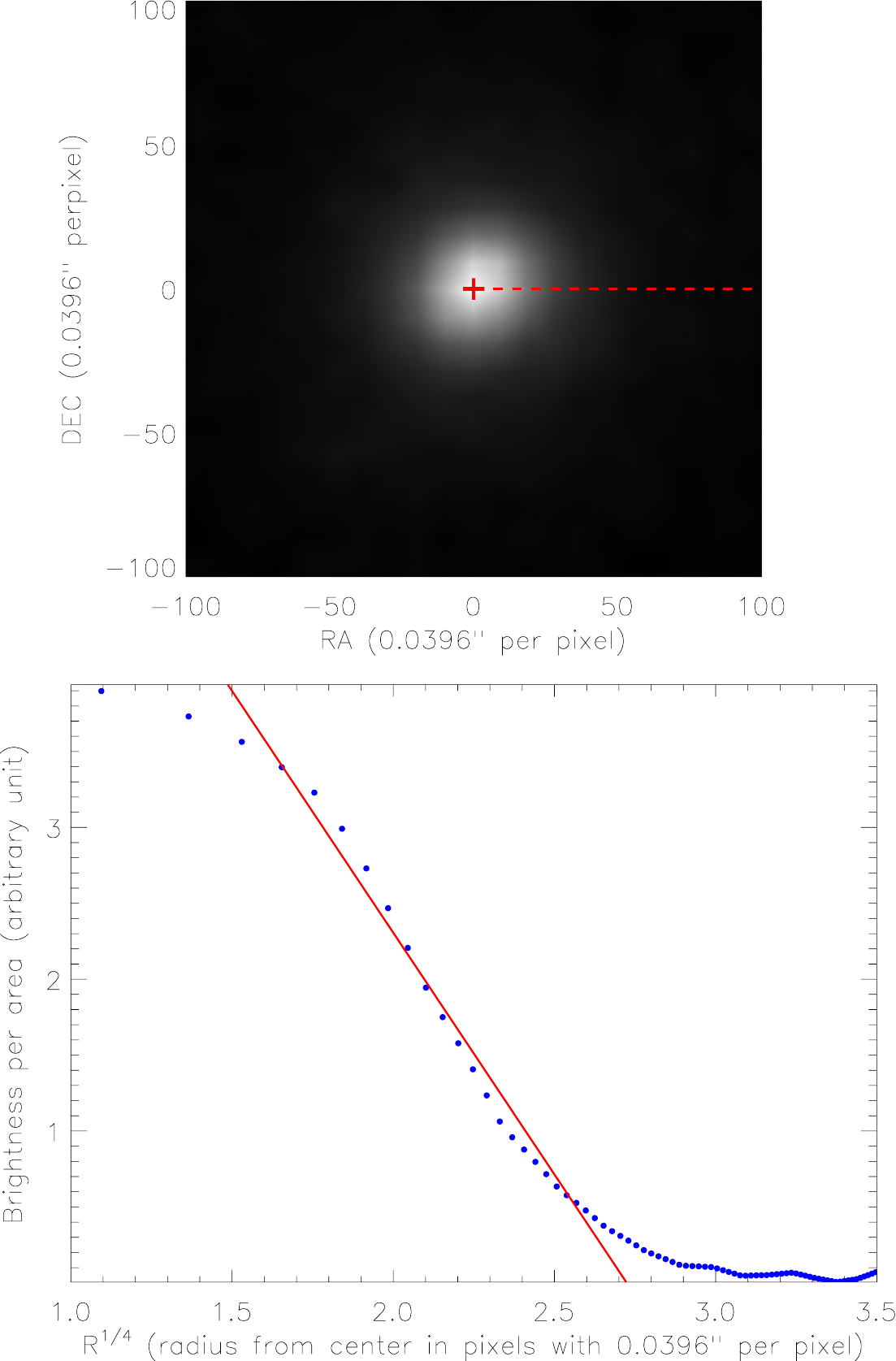}
\centering\includegraphics[width = 9cm,height=6cm]{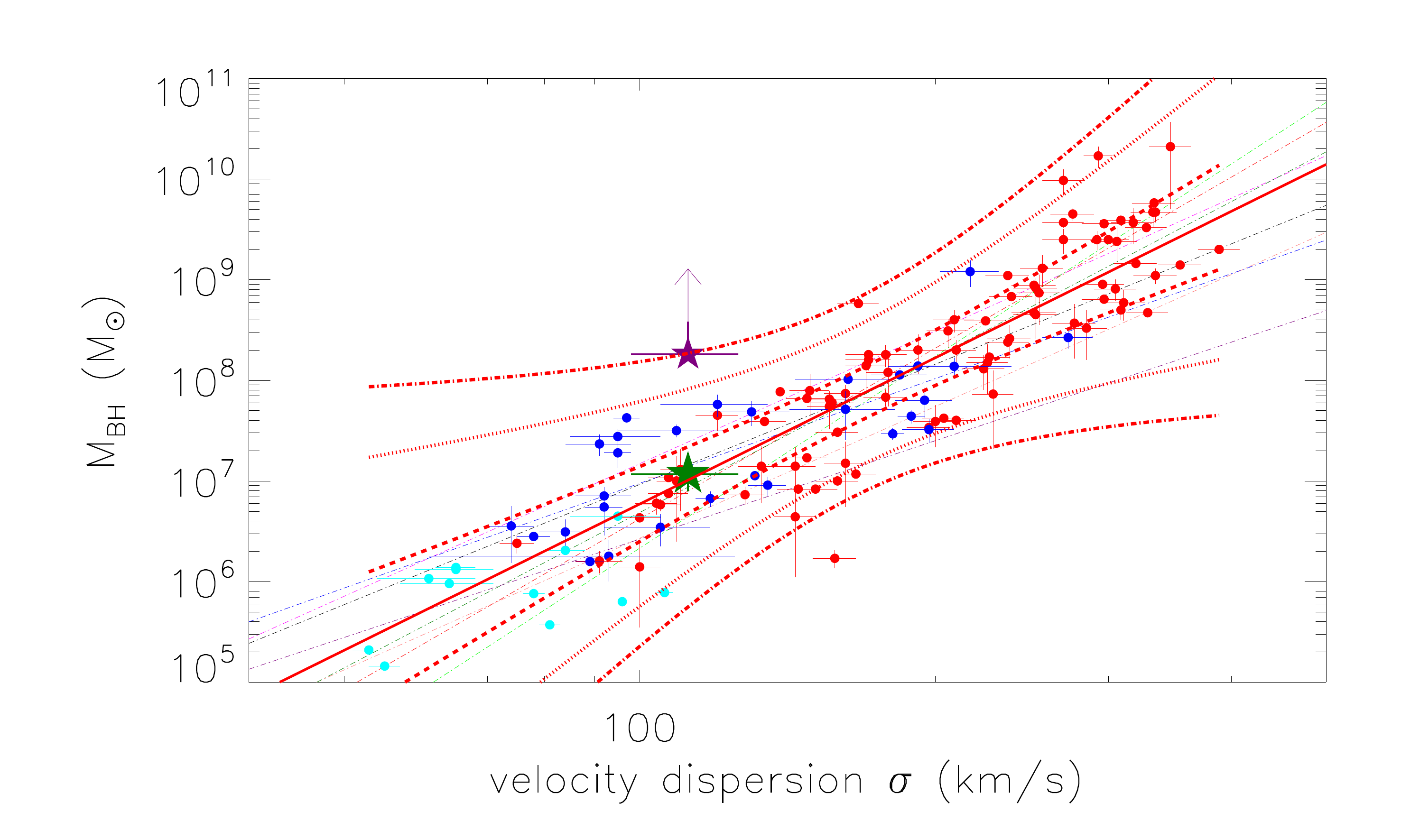}
\caption{Left panels: the photometric properties of SDSS J2334.  
The top left panel shows the $r$-band photometric image cut from the FITS (Flexible Image Transport System) image of the SDSS field, 
the red cross represents the peak value and the red dashed line represents the positions of obtaining the surface brightness profile. 
Here, the image is expanded from 10 $\times$ 10 pixels to 100 $\times$ 100 pixels using the linear interpolation method. 
The bottom left panel shows the surface brightness profile with horizontal coordinates of $pixel^{1/4}$, and the red solid line indicates the best-fitting 
results of using a linear function.
Right panel: on the correlation between stellar velocity dispersion measured through absorption features and BH mass of SDSS J2334. 
Solid five-point-star in dark green shows the BH mass of SDSS J2334 determined through applications of theoretical TDE model to 
long-term light curves. Solid five-point-star in purple shows the virial BH mass determined through assumed broad Balmer emission 
lines. Dot-dashed lines in green, in red, in magenta, in black, in pink, in purple, in blue and in dark green represent the 
relations through the quiescent galaxies in \citet{2015MNRAS.446.2330S}, in \citet{2013ApJ...764..184M}, in \citet{2013ARA&A..51..511K}, 
and through the RM (reverberation mapped) AGN in \citet{2015ApJ...801...38W}, the RM AGN with classical bulges in 
\citet{2014ApJ...789...17H}, the RM AGN with pseudobulges in \citet{2014ApJ...789...17H} and the RM AGN in \citet{2013ApJ...772...49W}, 
and through the TDEs in \citet{2021ApJ...907...77Z}, respectively. Solid circles in red, in blue and in pink show the values for 
the 89 quiescent galaxies from \citet{2015MNRAS.446.2330S}, the 29 RM AGN from \citet{2015ApJ...801...38W} and the 12 TDEs from 
\citet{2021ApJ...907...77Z}, respectively. Thick solid red line shows the best fitting results to all the objects: the 89 quiescent 
galaxies, the 29 RM AGN and the 12 TDEs, and thick dashed red lines, thick dotted red lines and thick dot-dashed red lines show 
the corresponding 3$\sigma$, 5$\sigma$ and 6$\sigma$ confidence bands to the best fitting results.}
\label{msig}
\end{figure}

	Based on the best descriptions to stellar absorption features, stellar velocity dispersion $\sigma_*$ is well measured 
as 112$\pm$14km/s, leading the central BH mass to be estimated about $10.2_{-4.5}^{+7.6}\times10^6{\rm M_\odot}$ as shown in 
Fig.~\ref{msig} through the well-known $M_{BH}-\sigma_*$ relation (between BH mass and stellar velocity dispersion) 
\citep{2000ApJ...539L...9F, 2000ApJ...539L..13G,2013ApJ...764..184M, 2013ARA&A..51..511K, 2017ApJ...838L..10B,2021ApJ...921...36B} 
widely accepted in quiescent galaxies \citep{2015MNRAS.446.2330S}, in active galaxies \citep{2015ApJ...801...38W} and also in 
TDEs \citep{2021ApJ...907...77Z}. The $M_{BH}-\sigma_*$ relation determined BH mass is well consistent with theoretical TDE model 
determined BH mass, providing further clues to support the central TDE and also support the SSP determined best descriptions in 
SDSS J2334. Here, in Fig.~\ref{msig}, the 89 quiescent galaxies from \citet{2015MNRAS.446.2330S} and the 29 reverberation mapped 
AGN from \citet{2015ApJ...801...38W} and the 12 TDEs from \citet{2021ApJ...907...77Z} are considered to draw the linear correlation 
between stellar velocity dispersion and BH mass through the Least Trimmed Squares robust technique \citep{2013MNRAS.432.1709C}
\begin{equation}
\log(\frac{M_{\rm BH}}{\rm10^6 M_\odot})=(2.22\pm0.04)+(4.83\pm0.22)\times\log(\frac{\sigma_*}{\rm 200km/s}).
\end{equation} 
And then the 3$\sigma$, 5$\sigma$ and 6$\sigma$ confidence bands to the linear correlation are determined and shown in the figure 
through the F-test technique. 

Data for galaxies of all morphological types are available to place on the $M_{\rm BH}$–$\sigma_*$ diagram. However, it is widely accepted that only massive early-type galaxies, particularly those formed predominantly through ‘dry’ mergers, are expected to follow $M_{\rm BH}$–$\sigma_*$ relation (e.g.  \citet{2008ApJ...688..159G,2013ApJ...775..116R,2018ApJ...863....1C,2020ApJ...898L...3B,2025arXiv250213202G,2013ARA&A..51..511K}). Therefore, further analysis of the photometric image properties of SDSS J2334 is needed. We have measured its surface brightness profile from the $r$-band image provided by SDSS to estimate the S{\'e}rsic index, as shown in the left panel of Figure \ref{msig}. The galaxy can be well described with the De Vaucouleurs profile (S{\'e}rsic index with n=4) indicated by the red line in the bottom left panel of Figure \ref{msig}. Considering possible AGN components near the central regions of the galaxy, the results from 0 to 0.2\arcsec of the surface brightness profile are not considered during the fitting procedure. 
Notably, the elliptical probability of SDSS J2334 provided by the SDSS Galaxy Zoo database (morphological classifications of galaxies \citep{2011MNRAS.410..166L}) is 0.633 with an nvote (the number of votes) of 30, indicating a favorable elliptical galaxy. Therefore, it is acceptable to estimate the black hole mass of SDSS J2334 using the $M_{BH}-\sigma_*$ relation.

\begin{figure*}
\centering\includegraphics[width=18cm,height=7cm]{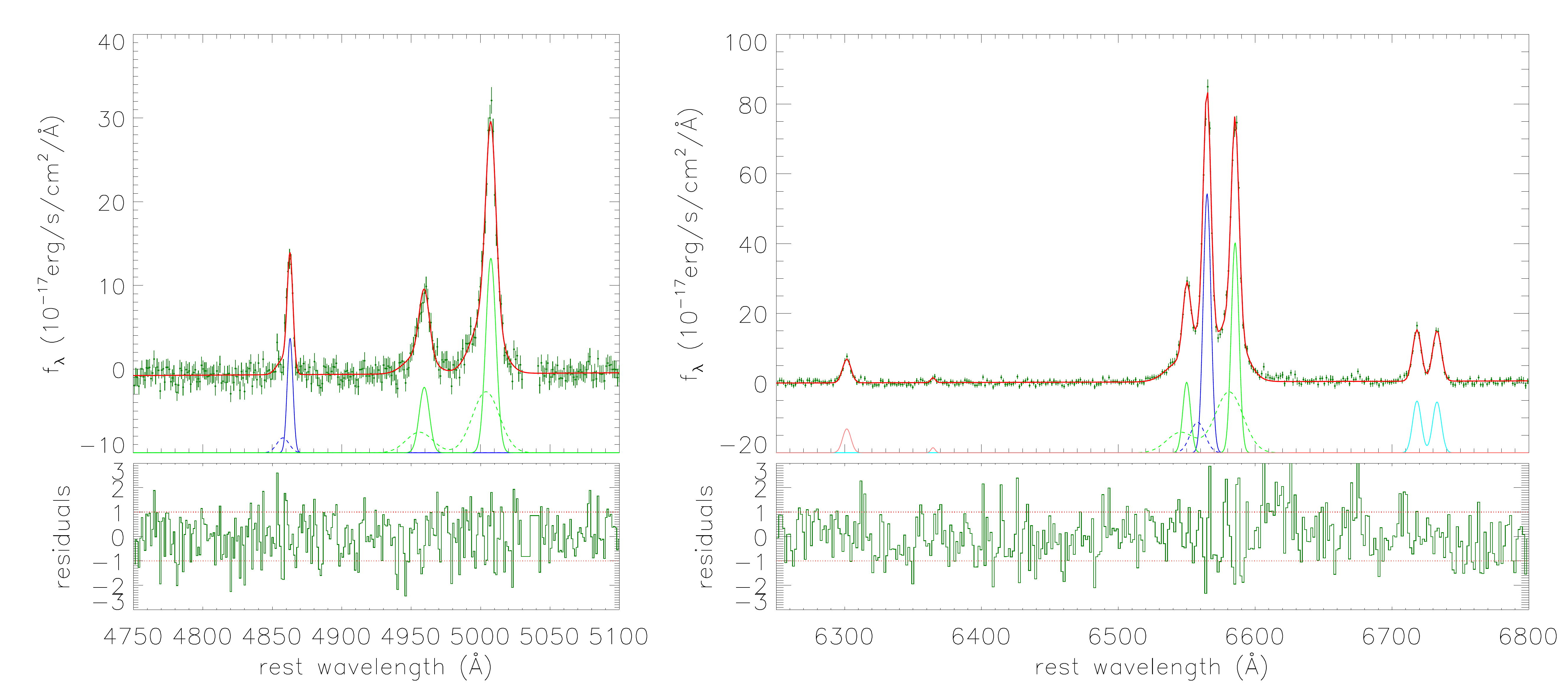}
\centering\includegraphics[width=18cm,height=7cm]{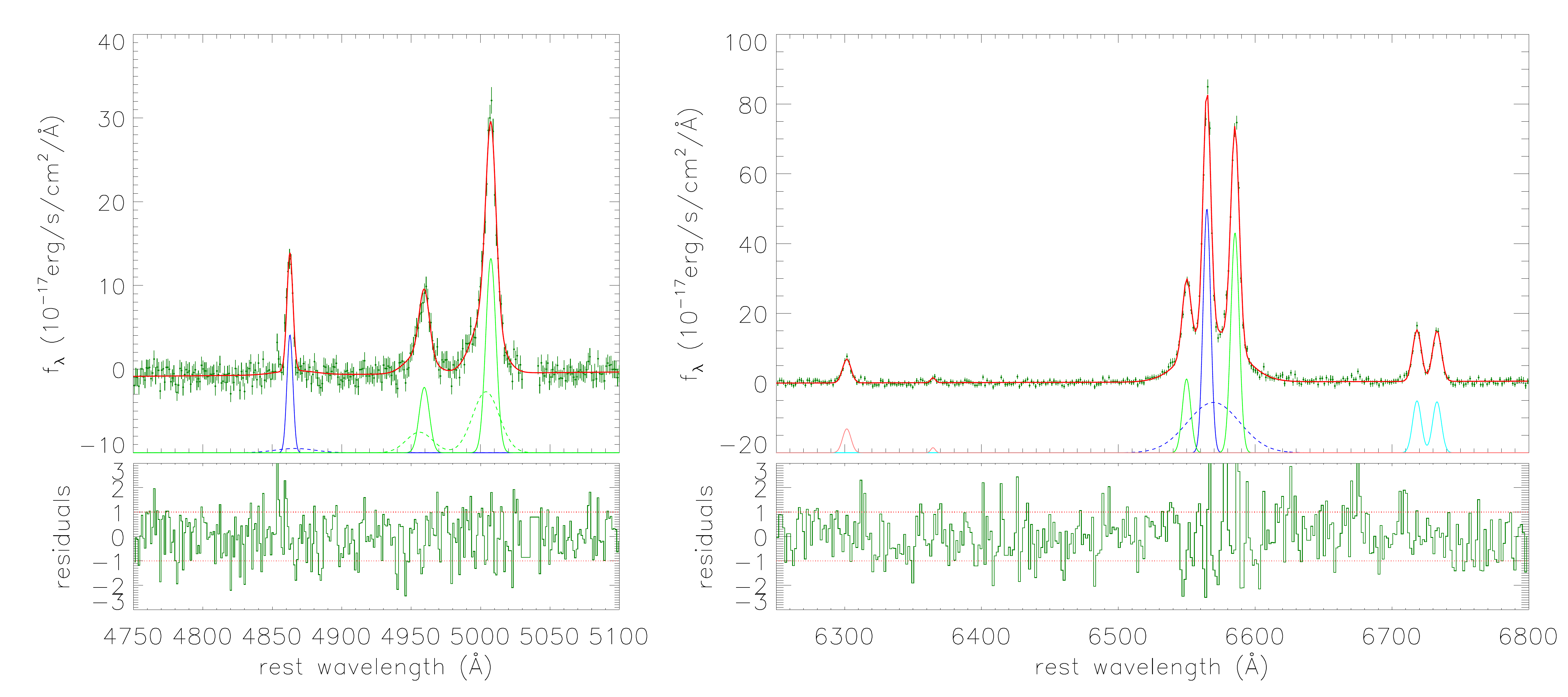}
\caption{Top panels show the best descriptions (solid line in red) and the corresponding residuals (bottom region) (line spectrum 
minus the best descriptions and then divided by uncertainties of SDSS spectrum) to the emission lines (small circles plus error 
bars in dark green) around H$\beta$ and around H$\alpha$ after considering blue-shifted wings in each narrow emission line.  
Bottom panels show the best descriptions (solid line in red) and the corresponding residuals (bottom region) to the emission 
lines (small circles plus error bars in dark green) around H$\beta$ and around H$\alpha$ after considering broad Balmer emission 
lines.}	
\label{line}
\end{figure*}

	After subtracting the host galaxy contributions and AGN continuum emissions (continuum luminosity at rest 5100\AA~ about 
$7.9\times10^{42}$erg/s), we can measure the emission lines around the H$\beta$ (from 4800\AA~to 5050\AA~) and H$\alpha$ (from 6480 
to 6800\AA) in the rest frame including emission lines of H$\beta$, H$\alpha$, [O~{\sc iii}]$\lambda4959,5007$\AA, 
[O~{\sc i}]$\lambda6300,6363$\AA, and [S~{\sc ii}]$\lambda6716,6731$\AA~ doublets, similar as what we have recently done in 
\citet{2021ApJ...909...16Z, 2022ApJ...937..105Z, 2024MNRAS.534L..23Z} and also as done in \citet{2005ApJ...627..721G, 
2011ApJS..194...45S}. Due to apparent blue wings in [O~{\sc iii}]$\lambda4959,5007$\AA~ doublet in SDSS J2334, there are two 
Gaussian functions applied to describe the core component and the blue wing of each narrow emission line. The best descriptions 
to the emission lines and corresponding residuals (calculated by line spectrum minus the best fitting results and then divided by 
uncertainties of SDSS spectrum) are shown in top panels of Fig.~\ref{line} with $\chi^2/dof\sim0.91$, through the Levenberg-Marquardt 
least-squares minimization technique. The measured line parameters of each emission line are listed in Table 1. Here, we should note 
that although extended components for blue wings are applied to describe the [O~{\sc i}] and [S~{\sc ii}] doublets, the measured 
fluxes of the extended components are around zero. Therefore, there are no parameters of extended components of the [O~{\sc i}] 
and [S~{\sc ii}] doublets in Table~1. Considering the determined second moments of the blue-shifted wings in the narrow emission 
lines, it is not necessary to consider a broad H$\alpha$ component.

	Moreover, besides two Gaussian components applied to describe each component in both [N~{\sc ii}] doublet and narrow 
H$\alpha$, three narrow Gaussian components plus a broad Gaussian component were applied to describe emission lines of [N~{\sc ii}] 
doublet and broad H$\alpha$ and narrow H$\alpha$, leading the corresponding best fitting results shown in bottom panels of 
Fig.~\ref{line} with determined $\chi^2/dof\sim0.97$. In order to confirm the fitting procedure determine broad component in 
H$\alpha$ not one broad component from normal central BLRs, the following three points can be clarified. First, considering the 
Virialization assumptions \citep{2002ApJ...571..733V, 2004ApJ...613..682P, 2005ApJ...630..122G, 2011ApJS..194...45S} to the broad 
component assumed from central normal BLRs, lower limit of virial BH mass of SDSS J2334 can be estimated as 
\begin{equation}
	M_{BH}=5.5\frac{R_{BLRs}\sigma^2_{H\alpha}}{G}~\sim~1.8\times10^8{\rm M_\odot}
\end{equation}
with $R_{BLRs}$ estimated by intrinsic continuum luminosity through the R-L empirical relation \citep{2013ApJ...767..149B} and  
$\sigma_{H\alpha}$ as the second moment of the determined broad H$\alpha$, after considering serious obscurations (estimated 
E(B-V)~2.2 through flux ratio of the determined broad H$\alpha$ to broad H$\beta$, relative to the theoretical value 3.1) on 
broad Balmer emission lines probably leading to lower line widths of broad emission components. The determined virial BH mass 
is at least one magnitude higher than the $M_{BH}-\sigma_*$ relation expected BH mass, providing clues to disfavor the 
determined broad component from normal central BLRs. Second, comparing with the fitting results by different model functions 
with considering broad Balmer components shown bottom panels of in Fig.~\ref{line} and with considering extended components 
related to radial flows shown in top panels in Fig.~\ref{line}, the F-test technique can be applied to confirm that the results 
shown in top panels in Fig.~\ref{line} should be preferred with confidence level higher than 6$\sigma$ (null hypothesis smaller 
than $10^{-10}$). Third, based on the measured continuum luminosity at 5100\AA~ $L_{con}\sim7.9\times10^{42}$erg/s in rest frame 
without considerations of any obscurations and the linear correlation between $L_{con}$ and the broad H$\alpha$ line 
luminosity for normal SDSS quasars as discussed in \citet{2005ApJ...630..122G}, the expected broad H$\alpha$ line luminosity 
is about 
\begin{equation}
	L_{H\alpha_b}~=~5.25\times10^{42}(\frac{L_{con}}{10^{44}{\rm erg/s}})^{1.157} {\rm erg/s}\sim2.7\times10^{41}{\rm erg/s}
\end{equation}
which is well consistent with the measured $L_{H\alpha_b}\sim2.1\times10^{41}{\rm erg/s}$ from the assumed broad component in 
H$\alpha$, indicating there were few effects of obscurations providing further clues to support that the determined broad H$\alpha$ 
component should be not the one from common central BLRs. Therefore, based on the discussed three points above, there are no 
reliable broad emission components from normal central BLRs in SDSS J2334.

\begin{table}
\caption{parameters for the emission lines}
\begin{center}
\setlength{\tabcolsep}{12pt}  
\renewcommand{\arraystretch}{1.2}
\begin{tabular}{cccc}
\hline\hline
 line  & $\lambda_0$ & $\sigma$ & flux \\
\hline
\multicolumn{4}{c}{Considering shifted wings in narrow emission lines, with $\chi^2/dof\sim575.32/631$} \\
	H$\beta_n$  & 4862.89$\pm$0.06	& 2.06$\pm$0.16	& 70.8$\pm$9.1 \\
	H$\beta_e$ &  4857.77$\pm$2.11	& 4.40$\pm$1.83	& 19.6$\pm$10.9 \\
	H$\alpha_n$ & 6564.91$\pm$0.09	& 2.81$\pm$0.11	& 523.6$\pm$52.3 \\
	H$\alpha_e$ & 6557.98$\pm$2.85	& 5.39$\pm$1.68	& 118.3$\pm$55.5\\
	\oc3 & 5007.48$\pm$0.14 & 3.69$\pm$0.19 & 214.5$\pm$20.1\\
	\oee & 5003.98$\pm$0.85 & 9.72$\pm$0.79 & 177.7$\pm$20.8\\
	\nii & 6585.39$\pm$0.06	& 2.82$\pm$0.07	& 426.1$\pm$15.2\\
	\nei & 6581.01$\pm$0.51 & 10.25$\pm$0.48 & 449.1$\pm$15.7\\
	\oi &  6301.61$\pm$0.22	& 3.07$\pm$0.25	& 52.7$\pm$3.8\\
	\oii & 6364.63$\pm$0.24	& 1.56$\pm$0.79	& 5.6$\pm$2.5\\
	\si & 6718.29$\pm$0.09	& 3.17$\pm$0.08	& 117.8$\pm$4.3\\
	\sii & 6732.91$\pm$0.09	& 3.17$\pm$0.09	& 115.9$\pm$4.1\\
\hline
	\multicolumn{4}{c}{Considering broad component in Balmer emission lines, with $\chi^2/dof\sim606.38/632$} \\
	H$\beta_b$ & 4866.39$\pm$0.34	  & 14.18$\pm$0.33  &	18.2$\pm$10.8 \\
	H$\beta_n$ & 4862.80$\pm$0.04  &	2.15$\pm$0.14  &	76.2$\pm$4.9\\
	H$\alpha_b$ & 6569.62$\pm$0.46  &	19.15$\pm$0.45  &	692.4$\pm$22.7\\
	H$\alpha_n$ & 6564.77$\pm$0.05  &	2.77$\pm$0.06  &	486.6$\pm$10.8\\
	\oc3 & 5007.49$\pm$0.14  &	3.68$\pm$0.19  &	214.2$\pm$20.2\\
	\oee & 5003.95$\pm$0.85  &	9.63$\pm$0.78  &	176.1$\pm$21.1\\
	\nii & 6585.32$\pm$0.05  &	3.09$\pm$0.05  &	491.1$\pm$9.7\\
	\oi & 6301.61$\pm$0.22  &	3.08$\pm$0.25  &	53.1$\pm$3.8\\
	\oii & 6364.63$\pm$0.24  &	1.63$\pm$0.81  &	5.9$\pm$2.5\\
	\si & 6718.29$\pm$0.09  &	3.19$\pm$0.08  &	  118.9$\pm$4.3\\
	\sii & 6732.91$\pm$0.09  &	3.19$\pm$0.08  &	117.1$\pm$4.1\\
\hline\hline
\end{tabular}\\
Notes: The first column shows which emission component is measured, H$\beta_n$ (H$\alpha_n$) means the narrow emission component, 
H$\beta_e$ (H$\alpha_e$) means the blue-shifted wing in Balmer line, H$\beta_b$ (H$\alpha_b$) means the broad emission component in 
Balmer line. [O~{\sc iii}]$\lambda5007$\AA~ and [O~{\sc iii}]$_e\lambda5007$\AA~ mean the core and the extended component 
in [O~{\sc iii}]$\lambda5007$\AA~ emission line. [N~{\sc ii}]$\lambda5007$\AA~ and [N~{\sc ii}]$_e\lambda5007$\AA~ mean the 
core and the extended component in [N~{\sc ii}]$\lambda6583$\AA~ emission line, when only shifted blue wings considered in each 
narrow emission line. The second, third and fourth columns show the central wavelength in unit of \AA, the second moment in unit 
of \AA~ and the emission flux in unit of $10^{-17}{\rm erg/s/cm^2/\textsc{\AA}}$, of each Gaussian emission component. 
\end{center}
\end{table}

\begin{figure*}
\centering\includegraphics[width = 15cm,height=5cm]{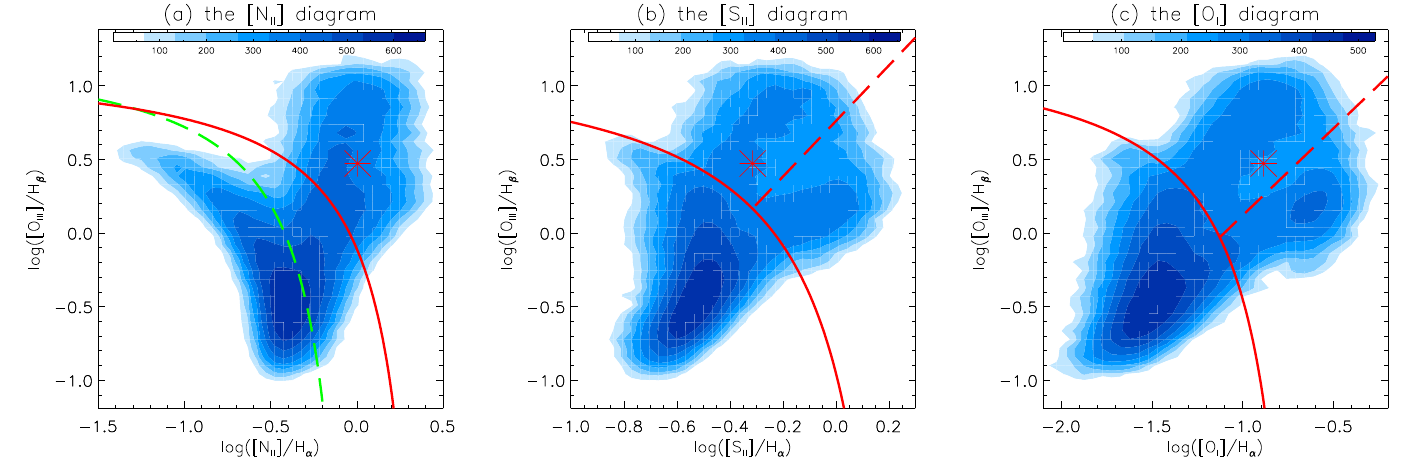}
\caption{SDSS J2334 (red asterisks) in the BPT diagrams of {[\mbox{O\,{\sc iii}}]/H$\beta$} versus {[\mbox{N\,{\sc ii}}]/H$\alpha$} 
(left panel), {[\mbox{O\,{\sc iii}}]/H$\beta$} versus {[\mbox{S\,{\sc ii}}]/H$\alpha$} (middle panel), and 
{[\mbox{O\,{\sc iii}}]/H$\beta$} versus {[\mbox{O\,{\sc i}}]/H$\alpha$} (right panel), where the contours filled by bluish colors 
represent properties of all narrow emission line objects collected from SDSS DR16, and solid red lines show the dividing lines 
between  \mbox{H\,{\sc ii}} galaxies and AGN reported in \citet{2001ApJ...556..121K}. In the left panel, The dashed green line 
and solid red line show the dividing lines between  \mbox{H\,{\sc ii}} galaxies and composite galaxies and AGN, reported in 
\citet{2003MNRAS.346.1055K, 2001ApJ...556..121K}. In the middle and right panels, dashed red lines show the dividing lines 
between Seyferts and LINERs reported in \citet{2006MNRAS.372..961K}.}
\label{BPT}
\end{figure*}

	Based on the measured line parameters after considering the blue shifted wing in each narrow emission line, the properties 
of SDSS J2334 are shown in Fig. \ref{BPT} in the BPT diagrams with emission line ratios calculated through the narrow components 
of the narrow emission lines. Based on the dividing lines \citep{2001ApJ...556..121K,2003MNRAS.346.1055K,2006MNRAS.372..961K} 
between AGN and composite galaxies and \mbox{H\,{\sc ii}} galaxies and LINERs in the BPT diagrams, SDSS J2334 is an AGN, not a 
LINER (low-ionization nuclear emission-line region) which could have probably different intrinsic physical ionization sources 
\citep{1996ApJS..102..161D,2011MNRAS.413.1687C} from BH accreting processes in AGN. Moreover, considering that the spectrum 
observation of SDSS J2334 occurred prior to the outburst, there are few effects of TDE emissions on narrow emission lines in 
spectrum of SDSS J2334. Therefore, based on the spectroscopic properties of SDSS J2334, there are power law continuum emissions, 
central AGN activities (not related to the central TDE) but no broad emission lines, indicating SDSS J2334 is a candidate of TT2AGN 
without hidden central BLRs. The emission lines in Figure \ref{line} top panel and the left bottom panel clearly 
reveal outflow components, which may originate from AGN activity or stellar winds. The primary difference between the two fitting 
methods is whether the H$\alpha$ and [N~{\sc ii}] fits include outflow components. The analysis of the same SDSS spectrum in the 
Reference Catalogue of Spectral Energy Distributions version 2 (RCSED2) project \footnote{https://rcsed2.voxastro.org} 
\citep{2017ApJS..228...14C} suggests that the broad H$\alpha$ can be explained by an outflow component in H$\alpha$ and 
[N~{\sc ii}], and this solution is preferred to a broad-line without an outflow component by F-test technique. In addition, 
RCSED2 also classifies SDSSJ2334 as an AGN based on the BPT diagram. Thus, the results from RCSED2 further support that our 
results are robust.

Before ending the section, one additional point should be noted. Although the F-test technique and the other methods are 
applied to accept the existence of the AGN continuum emissions and no apparent broad Balmer emission components included in the 
SDSS spectrum, the better choice to confirm such spectroscopic features is to do multi-epoch spectroscopic observations in the 
near future to check whether central AGN variability can lead to expected virialization expected variability of assumed broad 
Balmer emission components. More candidates like SDSS J2334 identified in the near future should provide further clues to support 
to to be against the proposed method in the manuscript. At the present stage, through the TDE model determined best descriptions 
to the photometric variability and the spectroscopic features of no apparent broad Balmer lines, the SDSS J2234 can be accepted 
as a potential candidate of TT2AGN.

\section{Disscussion}

\subsection{Mid-infrared/Radio data}
SDSS J2334 has been detected throughout the Wide-field Infrared Survey Explorer (WISE; \citet{2010AJ....140.1868W}). We have collected archival multiepoch mid-infrared (MIR) photometry from the ALLWISE\footnote{https://wise2.ipac.caltech.edu/docs/release/allwise/} and the most recent Near-Earth Object WISE (NEOWISE; \citet{2014ApJ...792...30M}) data releases. The ALLWISE data are only available for two observational epochs (MJD 55368 \& MJD 55549). NEOWISE provides a long-term light curve that spans the period from MJD 56645 (Dec. 2013) to MJD 60482 (Jun. 2024), with a half-year cadence. The light curves in W1 and W2 bands are presented in the middle panel of Figure~\ref{LC}.
The MIR light curves of SDSS J2334 show a slight rise from WISE, while the data from NEOWISE show a slow decay trend. Due to the large gap between WISE and NEOWISE, we are unable to conduct a further comparative analysis between the MIR and optical light curves. In the left panel of Figure \ref{msig}, the variation of WISE color (W1-W2) over time is displayed. The MIR color of SDSS J2334 is observed to turn red during the rise and blue during the decay,\footnote{Due to the large gap between WISE and NEOWISE, the highest flux data point may still be in the rising phase of the light curve.} possibly due to its optical light curves being characterized by a ``rapid rise, short peak, and long tail'' \citep{2025arXiv250310053Y}. In addition, the WISE color after the outburst (W1-W2 $\sim$ 0.65) is consistent with the typical value of AGN \citep{2018ApJ...866...26A}.



\begin{figure}
\centering\includegraphics[width = 7cm,height=5cm]{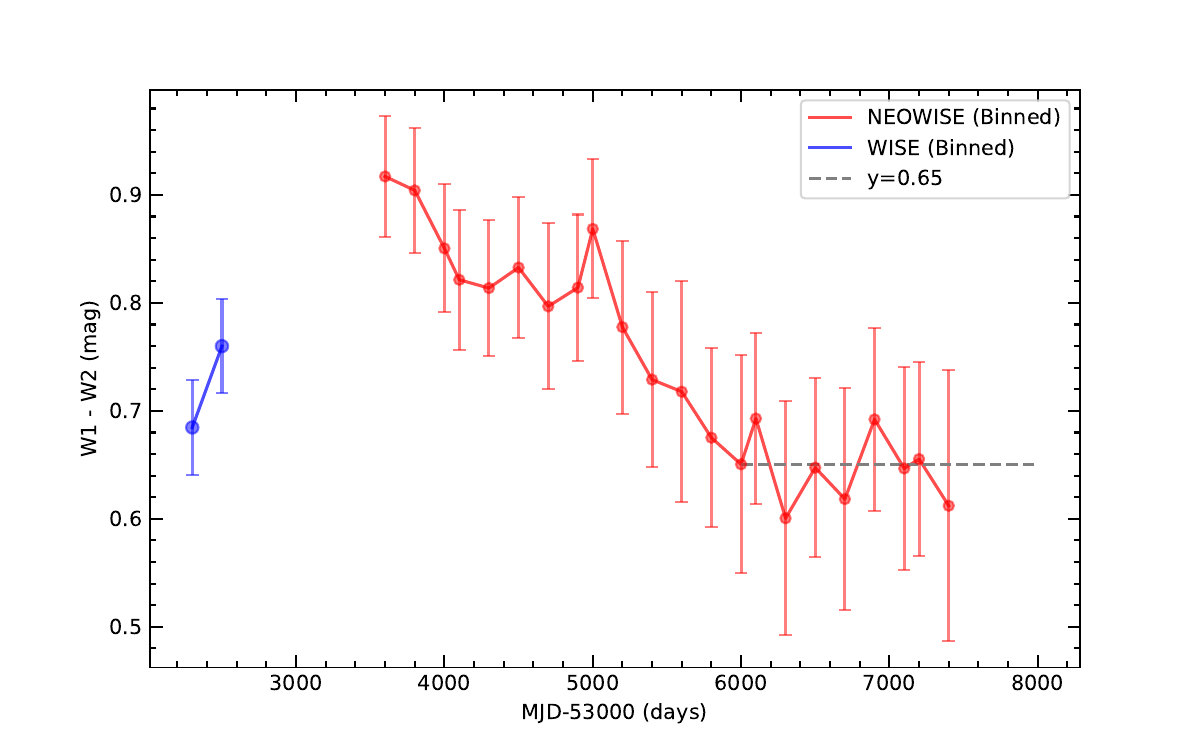}
\centering\includegraphics[width = 7cm,height=5cm]{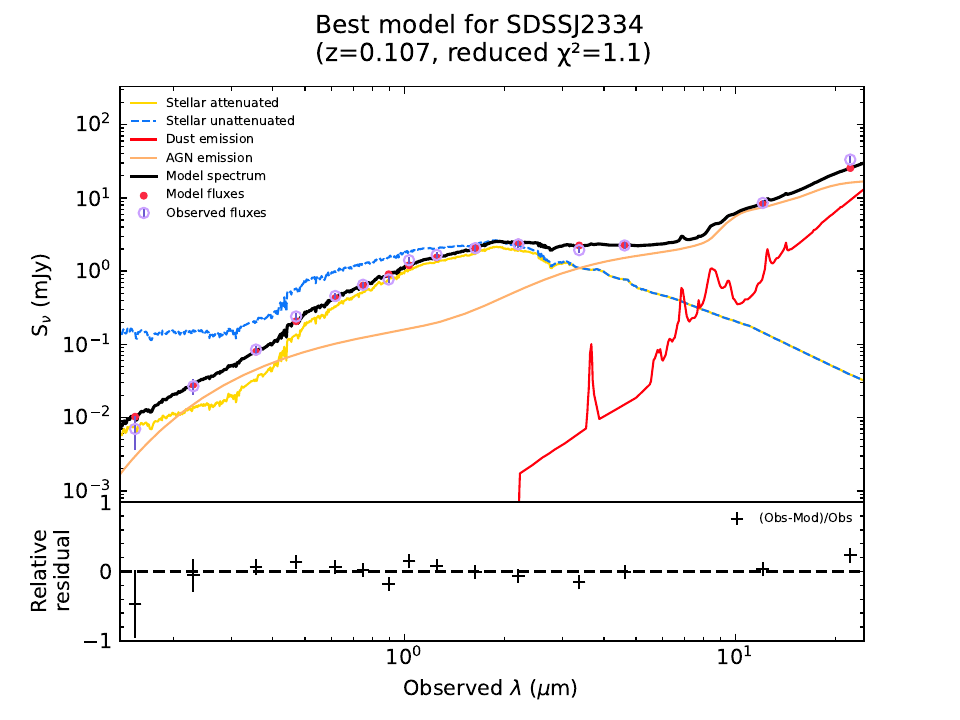}

\caption{Left panel: W1-W2 evolution of SDSS J2334, with binning applied to the data for each observation period of approximately six months. The gray dashed line represents the average value during the post-flare period. Right panel: SED fitting for SDSS J2334. The black line indicates
the best-fitting model. The blue, orange, red, and green lines represent
unattenuated stellar, attenuated stellar, dust, and AGN emission, respectively.
The lower panel indicates residual of the best fitting.} 
\label{color}
\end{figure}

We have also collected radio observation data reported by Very Large Array Sky Survey (VLASS; \citet{2020PASP..132c5001L} and Rapid ASKAP
Continuum Survey (RACS; \citet{2020PASA...37...48M}). VLASS observation at 2652 days after the optical peak (MJD=55381) gave 1.176±0.29 mJy at 3 GHz \citep{2021ApJS..255...30G}, while RACS at 3218 days gave 2.069±0.546 mJy at 888 MHz \citep{2021PASA...38...58H}. According to these data, the radio spectral index $\alpha = -0.46 \pm 0.28$ (assuming a power-law spectrum $S_\nu \propto \nu^{-\alpha}$) is derived, which is consistent with typical values found in core-dominated AGN \citep{2024ApJ...974..241A}.
Considering that the radio observations were conducted $\sim$ 8 years after the TDE optical peak, we suggest that the radio emission may originate from underlying AGN activity.


\subsection{SED FITTING RESULTS}
To better understand the properties between the host galaxy and AGN in SDSS J2334, we have used the SED fitting code \texttt{CIGALE}~v0.12.1 (\citet{2005MNRAS.360.1413B, 2009A&A...507.1793N, 2019A&A...622A.103B}). This code can estimate the physical properties of galaxies and/or AGN. The SED data for SDSS J2334 used in our analysis was obtained from the RCSED2 database. We have performed SED fitting using galaxy-only templates with a $\chi^2$/DOF value of 3.7 and using galaxy + AGN templates with a $\chi^2$/DOF value of 1.1. This shows a better fit quality when including the AGN component. The best fitting results are presented in the right panel of Figure \ref{color}. The results reveal a significant AGN contribution to the optical emission, indicating that the central region around the central BH of SDSS J2334 is in the direction of our sight.

\subsection{Alternative explanations for the flare observed in SDSS J2334}
The best descriptions to the optical flares can be accepted evidence to support a central TDE in the SDSS J2334. However, 
it is necessary to discuss the other explanations to the optical flares in SDSS J2334, such as the additional explanations in 
\citet{2017MNRAS.470.4112G,2017NatAs...1..865K}. Based on the asymmetric profile of the flare in SDSS J2334, the model on microlensing should be 
disfavoured. After considering the AGN-like spectroscopic features in the following section, the model on supernovae should be 
disfavored. 
If accepted the intriguing potential cause of the observed optical flare is a stellar mass binary black hole merger 
within the dense medium of the central accretion disc, or an AGN flare, the model still provides evidence that the central region around the 
central BH of the SDSS J2334 is in the direction of our sight, leading to the same clues as provided by the TDE model discussed above. Therefore, based on the TDE model determined best descriptions and the probable explanation by a stellar mass binary black 
hole merger within the central accretion disk (or instability of central accretion disk), we can accept that the central region 
around central BH of SDSS J2334 is in the direction of our sight.



\section{Conclusions}

	Candidates for True Type-2 AGN have been previously reported in the literature, not only in low luminosity AGN but also 
in high luminosity quasars. However, the extreme existence of True Type-2 AGN is still an open question. There are two main 
reasons leading to mis-classifications of True Type-2 AGN. First, weak optical broad emission lines were overwhelmed in 
spectroscopic noises. Second, central regions for optical broad emission lines were actually seriously obscured. Here, in SDSS 
J2334, BH mass properties can be applied to overcome the defect in overwhelmed intrinsic broad emission lines, and the detected 
TDE can be applied to overcome the defect in seriously obscured central broad line emission regions. In SDSS J2334, TDE properties 
can be applied to confirm central regions in our line-of-sigh, and meanwhile, spectroscopic properties can be applied to confirm 
no broad emission lines. Combining the TDE expected variability properties and the spectroscopic features with only narrow 
emission lines, the SDSS J2234 can be accepted as a potential candidate of True Type-2 AGN. The results provide clues on a new 
independent technique to identify True Type-2 AGN through unique properties of optical TDEs detected in Type-2 AGN.

\section*{Acknowledgements}
We gratefully acknowledge the anonymous referee for giving us
constructive comments and suggestions to greatly improve our
paper.
We sincerely thank Xin-zhe Wang, Hai-Ming Zhang, You-Dong Hu for useful discussion. This work is supported by the National Natural Science 
Foundation of China (grants NSFC-12173020, 12373014 and 12133003). Gu gratefully thank the kind financial support from the 
Innovation Project of Guangxi Graduate Education. The paper has made use of the code of TDEFIT \url{https://tde.space/tdefit/} 
which is a piece of open-source software written by James Guillochon for the purposes of model-fitting photometric light curves 
of tidal disruption events and also made use of the code of MOSFIT (Modular Open Source Fitter for Transients) 
\url{https://mosfit.readthedocs.io/} which is a Python 2.7/3.x package for fitting, sharing, and estimating the parameters 
of transients via user-contributed transient models. The paper has made use of the MCMC code 
\url{https://emcee.readthedocs.io/en/stable/index.html}.

\appendix

\section{Appendix A:}
The forced photometry results of SDSS J2334 are shown in Figure \ref{force}.

\renewcommand{\thefigure}{A}

\begin{figure*}
\centering\includegraphics[width = 8cm,height=5cm]{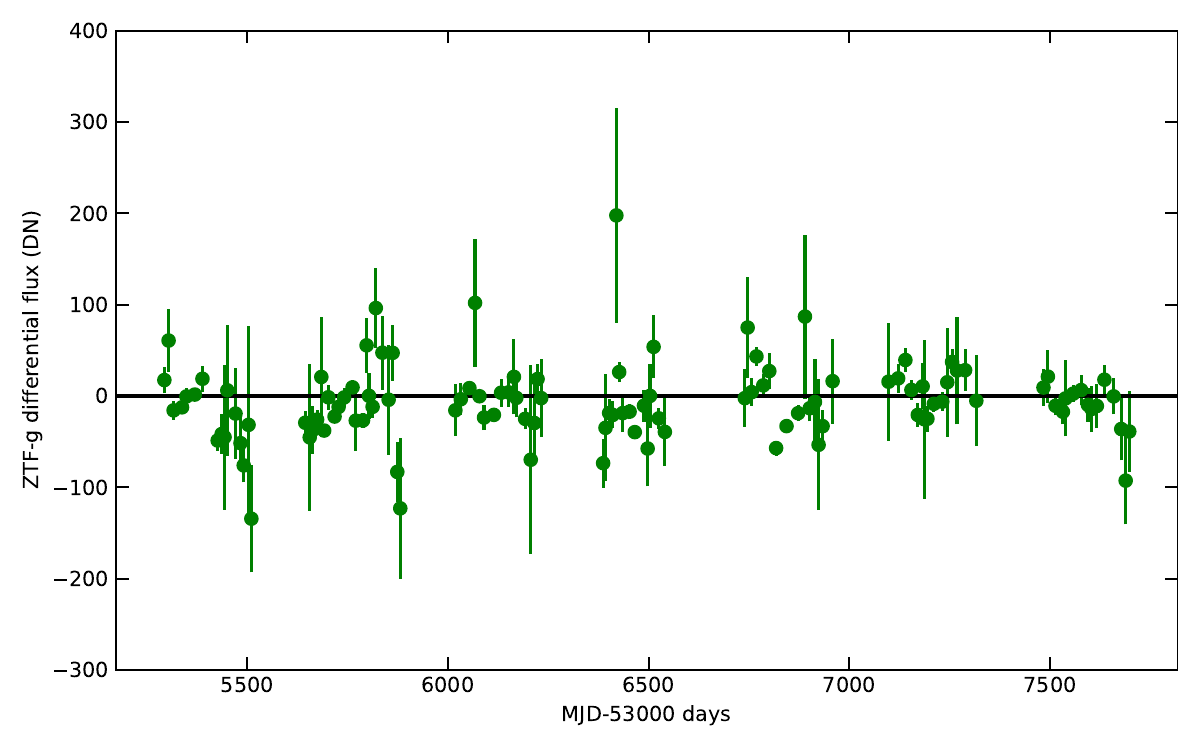}
\caption{Differential fluxes of SDSS J2334 in the ZTF-g filter were binned in groups of 15 consecutive data points, excluding large gaps to keep bins within continuous segments.} 
\label{force}
\end{figure*}

\bibliography{references}{ }
\bibliographystyle{aasjournalv7}

\end{document}